\documentclass{article}

\usepackage{PRIMEarxiv}

\usepackage[utf8]{inputenc} 
\usepackage[T1]{fontenc}    
\usepackage{hyperref}       
\usepackage{url}            
\usepackage{booktabs}       
\usepackage{amsfonts}       
\usepackage{nicefrac}       
\usepackage{microtype}      
\usepackage{lipsum}
\usepackage{fancyhdr}       
\usepackage{graphicx}       
\graphicspath{{media/}}     

\title{Towards Ethical Personal AI Applications: Practical Considerations for AI Assistants with Long-Term Memory
}

\author{
 Eunhae Lee \\
  Massachusetts Institute of Technology\\
  Cambridge, MA 02139 \\
  \texttt{eunhae@mit.edu} \\
}

\begin{document}
\maketitle

\begin{abstract}
One application area of long-term memory (LTM) capabilities with increasing traction is personal AI companions and assistants. With the ability to retain and contextualize past interactions and adapt to user preferences, personal AI companions and assistants promise a profound shift in how we interact with AI and are on track to become indispensable in personal and professional settings. However, this advancement introduces new challenges and vulnerabilities that require careful consideration regarding the deployment and widespread use of these systems. The goal of this paper is to explore the broader implications of building and deploying personal AI applications with LTM capabilities using a holistic evaluation approach. This will be done in three ways: 1) reviewing the technological underpinnings of LTM in Large Language Models, 2) surveying current personal AI companions and assistants, and 3) analyzing critical considerations and implications of deploying and using these applications.
\end{abstract}

\keywords{Personal AI \and AI Companions \and AI Assistants \and Long-term memory \and Artificial Intimacy}

\section{Introduction}
The remarkable progress of Large Language Models (LLMs) such as ChatGPT from OpenAI, Claude from Anthropic, and Gemini from Google has enabled human-like interactions through conversational interfaces. An active area of research is long-term memory (LTM), which allows these models to maintain context over extended periods and sessions, continuously learn about the user and their preferences, and effectively retrieve relevant information. 

One application area of LTM capabilities with increasing traction is personal (or personalized) AI companions and assistants. With the ability to retain and contextualize past interactions and adapt to user preferences, personal AI companions and assistants promise a profound shift in how we interact with AI and are on track to become indispensable in personal and professional settings. However, this advancement introduces new challenges and vulnerabilities that require careful consideration regarding the deployment and widespread use of these systems. 

The goal of this paper is to explore the broader implications of building and deploying personal AI applications with LTM capabilities using a holistic evaluation approach \cite{spector_data_2022}. This will be done in three ways: 1) reviewing the technological underpinnings of LTM in LLMs, 2) surveying current personal AI companions and assistants, and 3) analyzing critical considerations and implications of deploying and using these applications.

\section{Long-term memory mechanisms in AI}
\label{sec:longterm}

The evolution of LTM mechanisms in artificial intelligence has progressed from early symbolic systems to the sophisticated capabilities of contemporary LLMs. Traditional AI relied on symbolic methods like knowledge bases and rule-based systems, which stored and retrieved static information but lacked dynamic adaptability \cite{russell_artificial_2016}. As AI research advanced, neural network models emerged, showing promise in learning from data and generalizing to new situations. However, early neural models struggled with maintaining long-term context and adapting to user preferences over extended interactions, primarily using short-term memory. The introduction of Long Short-Term Memory (LSTM) and attention mechanisms partly addressed this issue. LSTMs were designed to tackle the vanishing gradient problem in earlier RNNs, allowing networks to retain information over more extended periods \cite{hochreiter_long_1997}. Attention mechanisms, particularly through Transformer architecture, further improved memory systems by enabling selective focus on relevant input data, enhancing the handling of long sequences \cite{vaswani2017attention}.

The advent of LLMs, like OpenAI's GPT series and Google's BERT, significantly advanced natural language processing. These models excel in translation, summarization, and text generation by leveraging large datasets and complex neural networks to produce coherent, context-aware outputs (Brown et al., 2020; Devlin et al., 2018). Despite these advances, attention mechanisms in LLMs face limitations in maintaining context and adapting over extended interactions, such as high computational costs, uneven information retention, and potential biases from training data. LLM memory lacks the depth and contextual recall of human long-term memory, which is crucial for applications like personalized recommendations or adaptive learning.

Key approaches to addressing these limitations to improve the LTM capabilities of LLMs include:

\begin{itemize}
    \item \textbf{Increased context length:} Research techniques, like sparse attention mechanisms and quantization, aim to increase context length without proportional computational requirements \cite{wang_beyond_2024}.
    \item \textbf{External knowledge base:} Retrieval-augmented generation (RAG) outsources memory functions to an external database, broadening the model's information access beyond its immediate training data \cite{lewis_retrieval-augmented_2021}.
    \item \textbf{Additional memory layers:} Memory-augmented neural networks (MANNs), such as Neural Turing Machines (NTMs) and Differentiable Neural Computers (DNCs), incorporate an external memory component for dynamic read and write operations \cite{graves_neural_2014, graves_hybrid_2016}.
    \item \textbf{Integrated memory:} MemoryBank, a long-term memory module, enables LLMs to store, recall, and update memories, adapting to user personalities by synthesizing past interactions using methods inspired by the Ebbinghaus Forgetting Curve theory \cite{zhong_memorybank_2023}.
\end{itemize}

\section{Case study: Personal AI companions}
Integrating long-term memory in personal AI systems can significantly enhance their functionality by enabling them to continuously learn from past interactions and adapt to user preferences over time, providing a deeply personalized experience. These models can be very powerful – for instance, AI companions can offer social companionship, solace to individuals in isolation or long-term care, and digital therapy that evolves to meet users' psychological needs \cite{chaturvedi_social_2023}. These models can also be trained to create digital twins that can serve as interactive avatars for celebrities, engaging fans in a personalized manner. Personal AI assistants can learn user preferences and effectively manage tasks with less human oversight. 

The addition of LTM capabilities in LLMs opens doors to future innovations in user experience design. Moreover, the introduction of new modalities such as voice-based or video-based models could tranform how users interact with AI. Interfaces can become more intuitive, allowing users to interact with these systems in increasingly natural and responsive ways.

Below is a brief survey of recent personal AI applications, as of May 2024. Note that these examples have been cherry-picked to be representative of each category and do not reflect the larger landscape of personal AI applications.

\begin{figure}
    \centering
    \includegraphics[width=0.9\linewidth]{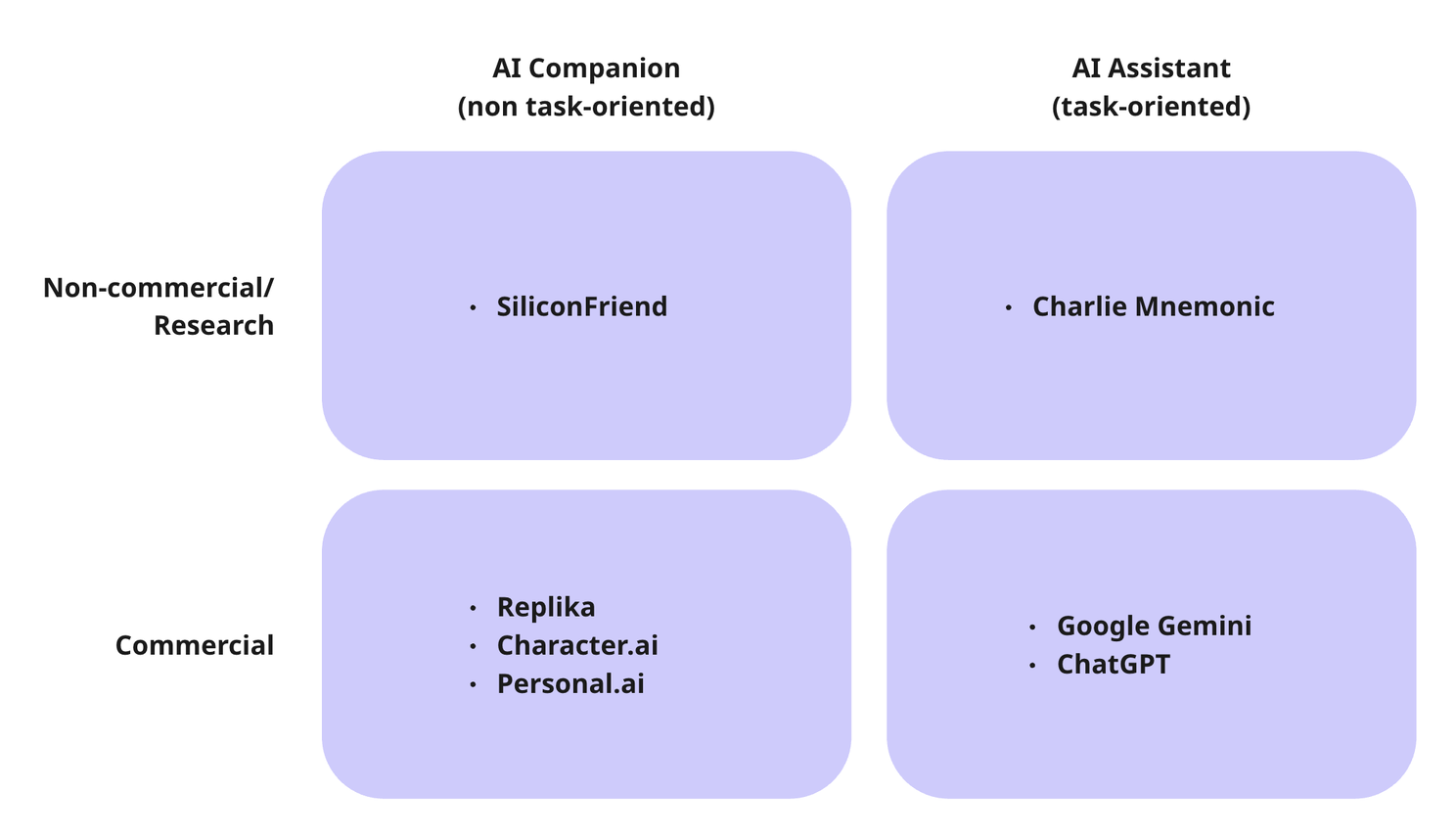}
    \caption{Brief taxonomy of personal AI applications (as of May 2024). Note that these examples have been cherry-picked to be representative of each category and do not reflect the larger landscape of personal AI applications.}
    \label{fig:personal_ai}
\end{figure}

\subsection{AI Companions}

Purpose: Provide emotional support, companionship, and personalized interactions.

\begin{itemize}
    \item \textbf{SiliconFriend:} Zhong et al. \cite{zhong_memorybank_2023} created SiliconFriend to evaluate MemoryBank's performance within LLMs. It focuses on long-term interactions, adaptability based on users' personalities, and meaningful emotional support through psychological data.
    \item \textbf{Replika: }Designed for deeply personal interactions, Replika helps users understand themselves better through empathetic, non-task-oriented interactions. While not (yet) equipped with LTM, the model continuously evolves to mirror the user's personality and enhance emotional engagement \cite{noauthor_replika_nodate}.
    \item \textbf{Personal.ai:} This platform creates personalized digital twins using unique models called Personal Language Models (PLMs). These models, equipped with long-term memory, allow for various personas, from professional assistants to companions. The platform is able to integrate new memories quickly and give users control over their data and the model's learning process \cite{personalai_differences_nodate}.
    \item \textbf{Character.ai:} enables users to create and interact with customized AI characters, providing personalized and engaging conversational experiences. It allows users to craft unique personalities and narratives, fostering creative and interactive storytelling. The platform continually adapts to user interactions, enhancing the realism and depth of character engagements \cite{characterai_characterai_nodate}.
\end{itemize}

\subsection{AI Assistants}
Purpose: Assist with tasks, increase productivity, and provide information.
\begin{itemize}
    \item \textbf{Charlie Mnemonic:} Touted as the first LLM-based personal assistant with long-term memory, Charlie Mnemonic uses GPT-4 to simulate human-like memory processes, offering personalized and enduring user interactions. It combines Long-Term Memory (LTM), Short-Term Memory (STM), and episodic memory to dynamically update memories and allow continuous learning without retraining \cite{goodai_introducing_2024}.
    \item \textbf{Google Gemini:} Formerly known as Bard, Google Gemini is an advanced AI assistant leveraging LTM for highly personalized experiences. It integrates across Google's ecosystem, adapting to user preferences to enhance functionality and engagement over time. It also has multimodal capabilities.
    \item \textbf{ChatGPT:} While not personalized, ChatGPT can remember and apply past interactions within a session to provide coherent, contextually relevant responses. Since February 2024, ChatGPT can remember (or forget) past conversations \cite{morris_chatgpt_2024}. It also has impressive multimodal features (GPT-4o).
\end{itemize}

While LTM is applicable for both AI companions and assistants, AI companions' focus on long-term personalized interaction strongly motivates the integration of LTM.

\section{Key considerations for designing and deploying personal AI companions}

\subsection{Tractable data}

The advancement of long-term memory mechanisms in LLMs, such as MemoryBank \cite{zhong_memorybank_2023}, can enable personal AI systems to efficiently store data from more extended interactions, update memory with user preferences and critical information, and integrate new knowledge with previous knowledge through abstraction. This approach necessitates meticulous data lifecycle management to ensure user privacy and security, including data acquisition, processing, storage, and use.

Data sparsity and scalability are key data-related challenges for AI companions and assistants. Data sparsity—having incomplete user data—can be mitigated as more data accumulates over time, improving personalization and user experience. Scalability is another challenge, as personal AI systems need to handle increasing data volumes without compromising performance or privacy. Strategies to improve scalability include expanding the context window of LLMs, utilizing cost-reducing computational techniques \cite{wang_beyond_2024}, and implementing forgetting mechanisms that mimic human memory \cite{zhong_memorybank_2023}. Additionally, innovative memory abstraction methods can reduce the need to store or retrieve extensive raw data while preserving the quality of generated content \cite{zhong_memorybank_2023, goodai_introducing_2024}.

\subsection{Data privacy and security}

Privacy must be a fundamental consideration when designing personal AI companions and assistants, as these systems process deeply personal and sensitive data. Balancing the benefits of deep personalization with the responsibility of protecting user data rights and privacy is critical for user trust and adoption. 

One area of privacy is user consent and autonomy, allowing users control over what data the AI collects and retains. Achieving high personalization often requires sensitive information, and while one may simply choose not to share, that would limit the personalization experience. Thus, it is essential to build robust privacy measures across the data lifecycle, including data collection, storage, access, confidentiality, and user management, including rights to edit or delete data. Applying technologies like federated learning and differential privacy to AI companions could enhance privacy by allowing data to be stored locally and adding noise to datasets, respectively \cite{spector_data_2022}.

Furthermore, data leakage and unauthorized third-party access are critical concerns for personal AI applications. Unlike human therapists, who are subject to strict confidentiality rules, AI therapists may not guarantee the same privacy \cite{hughes_artificial_2023}. Integrating AI assistants with other applications presents further challenges, such as when AI assistants are used within broader digital ecosystems (e.g., Google Gemini) and involve sharing personal data across apps. Preventing data leakage and strictly controlling third-party access is essential to maintaining user trust. Additionally, features like “like” or “dislike” buttons for AI-generated responses should be designed to respect privacy, ensuring feedback improves response quality without compromising individual context.

Data security is a crucial concern for personal AI companions and assistants due to their access to sensitive information, such as financial details, health records, and personal thoughts. For instance, chat models that can deduce behaviors and create a detailed digital twin of a person increase the risk of impersonation and misuse. These risks can be heightened by AI models with long-term memory and multimodal capabilities, which can be exploited if devices are lost or stolen. Additionally, public sharing of AI models, such as digital replicas of celebrities, can lead to exploitation for scams, adding another layer of security challenges. Ensuring robust protection against these vulnerabilities is essential to safeguarding user privacy and maintaining trust in these technologies.

\subsection{Resilience and understandability}

AI companions must be resilient to failures and robust against abuse due to their deeply personalized nature and the sensitive data they handle. Unexpected data loss can profoundly impact users, causing emotional distress, as seen in cases such as the abrupt shutdown of the AI chatbot Soulmate, which left many users distraught \cite{chayka_your_2023}. While data backups can help, they must be managed with stringent privacy measures.

For AI companions focused on emotional and social interactions, users may not need to understand the causal mechanisms behind responses as much as they do for critical decision-making systems, such as in medical settings. Additionally, reproducibility is challenging due to the highly personalized nature of these interactions. However, ensuring these systems are understandable and auditable when necessary is important, given their profound impact on individuals and society.

\subsection{Social, ethical, and legal implications}
\subsubsection{Lures and dangers of artificial intimacy}

The emotional dynamics between humans and AI companions are complex. While users generally understand they are interacting with machines, the emotional impact can be profound \cite{chayka_your_2023}. Instances of emotional distress, such as perceived relationship “breakups” with Replika, highlight the potential for emotional manipulation by AI \cite{brooks_i_2023}. As AI agents cannot inherently have human emotions, they can only simulate human empathy through computational algorithms \cite{liu-thompkins_artificial_2022}. Artificial empathy can inadvertently make vulnerable people feel manipulated by machines \cite{hughes_deep_2023}.

As AI companions become more sophisticated and capable of fostering deep emotional connections with users, the potential risk of users becoming overly reliant on them for social interactions and decision-making grows. Studies show that affective trust is as important as cognitive trust – the more they like an AI agent, the more likely they are to trust its outputs \cite{erdmann_understanding_2022, kyung_rationally_2022}. While appropriate reliance on AI can be beneficial, there are concerns that it may promote over-reliance and erode autonomy and critical thinking in the long term \cite{hughes_deep_2023}.

Artificial intimacy introduces additional societal implications, potentially altering how humans relate to each other. This dependency could shift perceptions of ideal human interactions, normalizing artificial intimacy. While AI can offer solace to the lonely or isolated, such as those with chronic illnesses, it risks diminishing the quality of human connections and lowering expectations of human intimacy \cite{center_for_humane_technology_esther_2023}. By removing the friction of regular human relationships, digital interactions might lower our tolerance for the natural discomforts of human relationships and diminish our ability to engage with people who challenge us. While AI companions can be used to increase self-validation and boost confidence, they risk creating self-echo chambers, similar to concerns raised about LLMs \cite{sharma_generative_2024}. Thus, AI companions should be seen as tools, not replacements for human interactions \cite{hughes_artificial_2023}.

\subsubsection{Potential for societal harm: Who is responsible?}

Incidents like Replika’s AI companion allegedly promoting a crime \cite{chayka_your_2023} and a man committing suicide after extensive conversations with an AI chatbot \cite{lovens_sans_2023} highlight the potential for negative societal impact as a result of forming deep emotional bonds and trust with AI companions. While some argue that AI companions could provide a “safe space” for people to be their authentic selves \cite{winter_ai_2023}, AI companions could potentially reinforce harmful or self-destructive behaviors in an untethered way, especially since these applications are primarily designed to empathize, not challenge, the users. While AI chatbots displaying misaligned behavior is not new (e.g., Microsoft’s AI bot, Tay in 2016 \cite{reese_why_2016}), the increased danger of personal AI companions that can emulate human empathy lies in the deep emotional bonds users may develop, making them more susceptible to misaligned behaviors.

These examples suggest that, unlike private self-talk, there is something fundamentally different about private interactions with AI in its potential for societal harm. This raises critical ethical and legal questions for stakeholders, including LLM creators, users, infrastructure providers, and regulatory entities, on who is held accountable. There is a need to think deeply about how to design and implement legal and societal guardrails to ensure these interactions do not exacerbate harm. 

\subsubsection{Equitable experience for all}
Fairness in AI companions involves ensuring equitable distribution of benefits and risks across all users, regardless of socio-economic status. Wealthier individuals often have better access to privacy-enhancing technologies and premium services, leading to disparities in privacy and the benefits of AI companions. To ensure fairness, policies and product designs must offer robust privacy protections and equitable access to advanced features and data security for all users, not just those who can afford them. Addressing these issues is crucial for fostering trust and inclusivity in AI technologies, ensuring that all users can benefit without compromising their privacy or well-being.

\section{Ethical AI companions: From principles to practice}

\subsection{Setting the right objectives}

Setting clear and ethical objectives is crucial when building and deploying personal AI companions, given their deep emotional impact on users and broader social and ethical implications. Objectives should be built on thoughtful consideration of the ideal human values to be embedded in the design and use of the technology, as these objectives significantly influence the downstream design and functionality \cite{friedman_value_2013, friedman_human_2002}.

When developing AI companions, the well-being of end-users and society should be the guiding principle. This means deprioritizing narrow commercial objectives like engagement maximization or user acquisition, and instead focusing on enhancing human flourishing, meaningful relationships, and positive societal contributions \cite{elliott_towards_2021}.

However, setting the right objectives can be challenging due to the inherent conflict between business interests and societal goals. The aim of maximizing profit often contrasts with enhancing long-term user well-being, similar to dilemmas faced by gaming companies with addiction \cite{cemiloglu_towards_2020}. Robust governance mechanisms, stakeholder engagement, organizational change, and public oversight can help ensure that AI companions truly serve the greater good.

Moreover, the right metrics need to be set to measure progress and drive action and resources. Traditional metrics around profit and reducing risk may not fully capture the nuanced impacts of AI companions on human well-being, including mental health, social connections, and personal growth \cite{chatila_ieee_2019}. 

\subsection{Building with human well-being in mind}
Translating ethical objectives into the actual design and functionality of AI companions requires operationalizing ethical principles around human and societal well-being \cite{chatila_ieee_2019}. Bridging the gap between principles and practice involves various challenges, including the complexity of AI’s impacts, diffusion of accountability for ethical consequences, the organizational division between technical and non-technical experts, and the misalignment across disciplines in how they frame and approach responsible AI \cite{schiff_principles_2020, khan_ai_2022}.

Among numerous responsible and ethical AI frameworks, comprehensive impact assessments such as the IEEE 7010 standard provide a broad yet flexible way for organizations to put ethical AI principles into practice and identify specific areas of improvement \cite{schiff_ieee_2020}. These frameworks help guide the stakeholders involved in shaping the AI system to consider the long-term impacts and risks of AI companions on human well-being, such as weakening social connections, over-reliance on technology, and diminished critical thinking.

As technologies continue to evolve, the impacts will also evolve and become more complex and unpredictable. To ensure that AI companions are aligned with ethical principles, continual monitoring and evaluation of the broader impacts of AI companions on users and society is important. These efforts should invite the participation of multiple groups of stakeholders, including developers, decision-makers, policymakers, and civil society, making it a participatory process \cite{schiff_principles_2020}. Importantly, these conversations about the ideal role of technology in society should involve diverse voices from civil society to ensure an equitable, inclusive, and sustainable future.

\subsection{The need for safeguards and regulatory framework}

Despite its rapid growth and deepening integration into personal lives, the burgeoning field of AI companionship remains notably unregulated. This lack of oversight is concerning, given the significant difference between using AI for mundane tasks and using it to simulate personal relationships. Unlike traditional social networks that facilitate human connections, AI chatbots offer a direct, albeit artificial, connection, often without the safeguards typically associated with human-to-human interaction. The current environment lacks clear benchmarks, standards, or guardrails, raising significant ethical questions about the nature of these interactions. Are they merely private thoughts, or do they represent a new form of social interaction that requires careful consideration and potential regulation? This debate brings privacy vs. security concerns to the forefront.

As personal AI systems become more integrated into our lives, timely and appropriate measures are necessary to safeguard personal data, ensure security, and promote human well-being. The EU AI Act prohibits certain AI practices considered an unacceptable risk, including exploiting people's vulnerabilities due to age, disability, or socio-economic status \cite{hainsdorf_2023}. The existing act could be expanded to include emotional manipulation of vulnerable individuals. Content moderation and user privacy controls can further mitigate exploitative AI practices targeting the vulnerabilities of users. Attention must also be paid to the content generated by these systems to avoid perpetuating biases or disseminating false or harmful information.

\subsection{Educating and empowering the public}

Educating the public about the potential harms and risks of AI companions should parallel highlighting their benefits and enjoyment. Balanced awareness ensures users are fully informed of the risks and can engage with these technologies responsibly, safeguarding their well-being. Engaging in dialogue, transparent communication from developers, public education efforts, and collaboration with advocacy groups can all contribute to a more informed user base.

Moreover, there is a need for stakeholders to better understand how people interact with, relate to, and depend on AI, especially as they become more sophisticated and human-like. Research in Human-AI interaction, HCI, Human-Robot interaction have been looking into the intricacies of how people interact with these systems. Lessons can also be learned from experts in human relationships and psychology–for instance, psychotherapists and social workers–who could speak deeply about how these systems are impacting the psychological and social aspects of people’s lives \cite{center_for_humane_technology_esther_2023}.

\section{Conclusion}
With the rapid evolution of technology such as long-term memory in AI systems, personal AI companions and assistants will become even more powerful and widespread in the near future. Emotional connections with AI systems can lead to heightened, sometimes unwarranted trust, making users vulnerable to manipulation. The key question remains: as technology evolves, how can it be used in beneficial ways without being manipulative? This paper has demonstrated the importance of a holistic evaluation of building and deploying personal AI companions by comprehensively reviewing technical challenges, privacy and security concerns, and broader societal implications. It highlights the need to set clear, humanity-centered objectives that consider long-term impacts and prioritize human well-being, operationalize these goals by involving multiple stakeholder groups, create effective and timely safeguards, and increase efforts to better inform the public about the benefits and risks of these applications. As personal AI companions and assistants become more deeply integrated into our daily lives, we must do so intentionally and with proper foresight to maximize the benefits and minimize the risks for everyone.


\bibliographystyle{unsrt}  
\bibliography{references}

\end{document}